\documentclass[12pt]{article}

\usepackage[margin=1in]{geometry}
\usepackage{amsmath,amssymb,graphicx,color}
\usepackage[titletoc,title]{appendix}
\usepackage{bm}
\usepackage{mathrsfs}
\usepackage{authblk}

    \title{Identification of relevant performance indicators in round-robin tournaments}
 \author{Andreas Heuer}
 \affil{University Muenster, Institute of Physical Chemistry, 48149 Muenster, Germany}

\begin{document}
\maketitle

  \begin{abstract}
 A myriad of different data are generated to characterize a
soccer match. Here we discuss which performance indicators
are particularly helpful to forecast the future results of a team via an estimation of the underlying team strengths with minimum statistical uncertainty.  We
introduce an appropriate  statistical framework and exemplify it for different performance indicators for the German premier soccer league. Two aspects are involved: (i) It is quantified how well the estimation process would work if no
statistical noise due to finite information is present. The related score directly expresses to which degree the chosen performance indicator reflects the underlying team strength. (ii) Additionally, the reduction of the forecasting quality due to statistical noise is determined. From both pieces of information a normalized value can be constructed which is a direct measure of the overall forecasting quality. It turns out that the
so-called packing rate works best. New perspectives of performance indicators enter when trying to understand the
outcome of single matches based on match-specific observations from the same match.  Implications for the purpose of forecasting as well as consequences for the interpretation of team strengths are discussed.
\end{abstract}

\section{Introduction}

With the help of statistical modelling new insight can be gained
about the underlying properties of soccer matches. A typical
example is the distribution of goals, since the
distribution of match results can be, to a good approximation,
taken as the outcome of two independent Poisson processes
\cite{Maher82,janke1,janke2}.  Actually, a detailed analysis reveals that the
actual number of draws is slightly larger than expected from the Poission distribution. This effect can be interpreted in psychological terms \cite{Riedl15}.

A Poisson distribution results if the
process of scoring a goal can be described as a probabilistic process, i.e.
as a probability per minute. Just like rolling a dice, the actual number of goals after 90 minutes can be quite different, despite identical probabilites. Within this work, this effect gives rise to a {\it random} contribution, acting as a kind of noise. Similarly, after, e.g., rolling the dice only for 12 times it is hard to find out that
each side has the identical probability of 1/6.

As a quantitative basis of our analysis we start with the definition of the team strength $S_A$ of a team $A$. It is given by the expected goal difference when team $A$ is playing against all other teams. The effect of the home advantage has to be taken out, see also below. One might naturally expect that the team strength is estimated based on the previous results of team $A$, see, e.g., \cite{Lee97,Dixon97,Dixon98,Rue00}.  As intensively discussed in this work, the estimation of $S_a$  may be also based on different performance indicators rather than on previous goals.
As outlined in \cite{Heuer10}, one can formally express the expected result of a match A vs. B on the basis of the team strengths $S_A$ and $S_B$.

Unfortunately, the number of goals is problematic from a
statistical perspective due to its small value and the
corresponding large random contributions. This results in a
somewhat limited relevance for forecasting purposes
\cite{Link16} and relatively frequent wins by the underdog \cite{ben3}. The same holds
for taking the ranks in the table as a measure, albeit some interesting time evolution of the ranks can be formulated, reflecting the nature of the distribution of team strenghts \cite{Morales}.

Thus, alternative measures have been taken into
account such as shots on goals, chances for goals, number of
passes, ball possession, or score box possessions for further
improvement \cite{Link16,Tenga10, Memmert17,heuer_buch}. It could
be shown on a quantitative level \cite{Heuer14} that chances for
goals are more powerful than the actual goals to forecast the
actual team strength. It turns out that for most variables it is
best to take the relative properties into account  (e.g. passes of team under consideration minus the passes by the opponents \cite{Pappalardo17}. A general review about different approaches can be found in \cite{Sarmento14}.

A new dimension in the data analysis has emerged due to the
availability of tracking data \cite{Pappalardo18}. A priori it is not evident how
relevant indicators can be extracted from these data. For example,
in \cite{Link16} the match dominance has been introduced which can
be either determined for a total match or even for a short period
of a match. It could be shown that the value of the match
dominance in a match displays a significant correlation with the
win probability, determined from the odds of bookmakers. Furthermore, the
availability of tracking data also allows one to perform a precise analysis on the level of the individual players; see, e.g., \cite{Liu16}.

For a long time the properties of passes and a related indicators,
the ball possession, have been of interest to sports researchers.
It turns out that the number of passes \cite{heuer_buch, Pappalardo17} as well
as the ball possession \cite{Link16} somewhat reflect the team
strength. The additional information from tracking data can be
used, e.g., to gain new insight about the behavior and possession
strategies of high and low percentage ball possession teams
\cite{Bradley14}. Furthermore, it may be used to refine the
definition of passes by using appropriately weighting factors.
This may help to identify truly successful passing behaviour. In
this spirit, Memmert and coworkers \cite{Memmert17}  have
additionaly taken into account the number of outplayed defenders.
It could be shown that the number of goals, scored during a game,
is significantly correlated with the number of outplayed
defenders. As further described below, these ideas gave rise to the so-called packing rate.

In this work we use a
theoretical framework which allows one to quantify how well a specific performance indicator $X$, e.g. the number of passes, is able to estimate the team strength. We remind again that knowledge of the team strength allows one to predict the expected future {\it goals} in a match.
Within this general approach different
questions can be posed and answered: (1) To which degree does the performance indicator
$X$ reflect the team strength in the limit that $X$ is determined
on the basis on an (hypothetical) infinite number of previous
matches?  In this limit the disturbing effect of noise is absent.
This question is of key relevance when defining new indicators,
e.g. on the basis of the tracking data. Within the present
approach it can be checked whether this variable reflects a more
or less irrelevant feature of a team or whether it is at the heart
of winning matches. (2) To which degree is the information about $X$, taken from a {\it finite} number of previous matches, reduced?
It is exactly this noise aspect which is expected to be
problematic when estimating the team strength based on the previous goal difference. (3) How do (1) and (2) connect to formulated an overall criterion for the forecasting quality of a performance indicator $X$.  (4) So far, the information from previous
matches is taken to forecast the future outcome. New insight is
gained if the performance indicator $X$ and the goals are taken from the same
match. Actually, this question was discussed in the context of the
7:1 result of the match Germany vs. Brazil during the world
championship 2014. Most performance indicators (e.g. ball possession, shots on
target, corners) spoke in favor of Brazil and did not contain any
indication for this extreme result \cite{internet1}.
Interestingly, a significant asymmetry was present for the impect
rate (Germany: 84; Brazil: 53) \cite{internet2}. Indeed, it will
turn out below that in a broader sense this performance indicator is indicative
about the actual result.

The outline of this work is as follows. In Section 2 we formulate
the statistical framework within which the data are analysed. In
Section 3 the different performance indicators are introduced. Section 4
contains the results of the statistical analysis. We conclude
in Section 5 with a discussion and a summary. Please note that the formalism is in principle applicable to many types of sport data (in particular to round-robin tournaments) so that the application to soccer should be mainly considered as a working example.

\section{Modelling approach}

\subsection{Statistical framework: correction for attenuation}

We start by introducing the framework for the {\it correction for attenuation} which
is a well-established procedure mainly in the psychological
literature \cite{Spearman04,Muchinsky96} but also has applications
in different fields such as physics; see, e.g., \cite{Kinahan98}. We first formulate it in general terms and then, in the next subsection, translate to the specific case of a sports league (here: soccer) where each team plays twice against any other team. We consider 2N variables $x_i$ and $y_i$ with $i \in {1,...,N}$.
During the formal definitions we assume that the distribution of
the set $\{x_i\} $ has vanishing average value and is
characterized by the variance $\sigma^2(x)$. Generalizations to
non-vanishing averages are straightforward.  Analogous properties
are found for the set $\{y_i\}$, characterized by the variance
$\sigma^2(y)$. All $x_i$ are measured twice, yielding the measurements
$x_i^{(1)}$ and $x_i^{(2)}$. Each measurement is characterized by
measurement noise with variance $\epsilon^2(x)$. Similarly one has
the measurements $y_i^{(1)}$ and $y_i^{(2)}$ and the measurement noise
$\epsilon^2(y)$. Now we define the Pearson correlation coefficients
$r(x^{(1)},y^{(2)}),r(x^{(1)},x^{(2)}) $, and
$r(y^{(1)},y^{(2)})$ . By definition one has
\begin{equation}
\label{corrxy}
r(x^{(1)},y^{(2)}) = \frac{cov(x^{(1)},y^{(2)})}{\sqrt{\sigma^2(x) + \epsilon^2(x)}\sqrt{\sigma^2(y) + \epsilon^2(y)} }.
\end{equation}
with $cov(x,y) = (1/N) \sum_i x_i y_i$
and
\begin{equation}
\label{corrxx}  r(x^{(1)},x^{(2)}) = \frac{\sigma^2(x)}{\sigma^2(x) +
\epsilon^2(x)} = \frac{1}{1+\epsilon^2(x)/\sigma^2(x)}.
\end{equation}
An analogous equation holds for $r(y^{(1)},y^{(2)})$.  This allows us to rewrite Eq.\ref{corrxy} as
\begin{equation}
\label{corrxy2}
r(x^{(1)},y^{(2)}) = \frac{cov(x^{(1)},y^{(2)})}{\sqrt{\sigma^2(x) \sigma^2(y)}}\sqrt{r(x^{(1)},x^{(2)})}\sqrt{r(y^{(1)},y^{(2)})}.
\end{equation}
The first term on the right side can be interpreted as the
correlation of $x^{(1)}$ and $y^{(2)}$ if no noise were present.
Therefore, we abbreviate this term as
$r_{ideal}(x^{(1)},y^{(2)})$, finally yielding
\begin{equation}
\label{corrxy3}
r(x^{(1)},y^{(2)}) = r_{ideal}(x^{(1)},y^{(2)})\sqrt{r(x^{(1)},x^{(2)})}\sqrt{r(y^{(1)},y^{(2)})}.
\end{equation}
From now one we omit the indices $(1)$ and $(2)$ for the sake of brevity.

Eq.\ref{corrxy3} has a straightforward interpretation. The
observed correlation $r(x,y)$ is smaller than unity for two
different reasons. First, even without noise the correlation
between the variables $x$ and $y$, in general, is not perfect.
This is expressed by $r_{ideal}(x,y) < 1$. Second, a further
reduction occurs due to the measurement noise. More specifically,
the reduction depends on the ratios $(\epsilon^2(x)/ \sigma^2(x))$
and $(\epsilon^2(y) / \sigma^2(y))$. Therefore, this reduction
effect is prominent if the noise variance is large as compared to
the variance of the distributions of the $x_i$ and $y_i$,
respectively, as expressed by Eq. \label{corrxy2}.

\subsection{Application to a soccer league}

Our specific goal is to forecast the goal difference
$G_{\Delta,i}^{(2)}$ of a team $i$ in the second half of the
season based on some performance indicator $X_i^{(1)}$, recorded
for that team in the first half. This predicted goal difference is identical to
the team strength of that team and in particular allows one to formulate an unbiased forecast of individual matches \cite{Heuer10}. The number of teams in a league is denoted $N$.

All performance indicators,
discussed in this work, are defined as an average per match. Since
the team plays against all other teams exactly once in each half,
there is no bias due to the choice of the opponents. A natural
choice for the performance indicator would be the goal difference
as well but other choices are also possible and will turn out to be
superior.

Of key relevance is the Pearson correlation  $r(X^{(1)},
G_\Delta^{(2)})$. It is a measure for the forecasting quality. A perfect
forecasting would be reflected by a perfect correlation of unity.
For a closer discussion of this correlation we use
Eq.\ref{corrxy3} to rewrite $r(X^{(1)}, G_\Delta^{(2)})$ as
\begin{equation}
\label{corrxy4} r(X, G_\Delta) = r_{ideal}(X,
G_\Delta)\sqrt{r(X,X)}\sqrt{r( G_\Delta, G_\Delta)}.
\end{equation}
Again we omit the indices, related to the first and the second
half of the season. The key quantity is $r_{ideal}(X, G_\Delta)$. Since it denotes the
correlation of $X$ and $G_\Delta$ in the absence of random
effects, it expresses to which degree the performance indicator $X$
reflects the team strength $S$. The random non-deterministic aspects
of a soccer match result in non-perfect correlations, i.e.
$r(X,X)< 1$ and  $r(G_\Delta , G_\Delta ) < 1$. Thus, as
expected both factors reduce the quality of the actual forecasting
of $G_\Delta $ based on $X$.

It is still problematic that $r(X , G_\Delta )$ also contains the effect of the $r(G_\Delta , G_\Delta ) < 1$, i.e. the measurement noise of the observable which we want to forecast. This measurement noise, however, is not relevant for the quality of the performance indicator $X$. Therefore, we finally introduce $A(X)$ via
Eq.\ref{corrxy4} as
\begin{equation}
\label{corrxy5} A(X) \equiv \frac{r(X, G_\Delta
)}{\sqrt{r(G_\Delta, G_\Delta)}} = r_{ideal}(X, G_\Delta
)\sqrt{r(X,X)}.
\end{equation}

$A(X)$ is an objective measure how well the performance indicator
$X$ is able to forecast the team strength, i.e. its goal difference,  in the second half of the season. The maximum value of $A(X)$  is unity. Since it is independent of the measurement noise of the goal difference of the second half of the season, the situation $A(X) = 1$ reflects a mathematically optimum forecast and, thus, can be taken as an unbiased estimation of the quality of the performance indicator $X$.

From Eq. \ref{corrxy5} we can see that a high forecasting quality (i.e., a large value of $A(X)$) requires, first,  that the performance indicator contains very similar (ideally identical) information as the team strength and, second, the actual realization of $X$ should be as noise-free as
possible.  In practice, we determine the three
correlation coefficients $r(X, G_\Delta), r(G_\Delta, G_\Delta)$
and $r(X,X)$ from the actual data and obtain $r_{ideal}(X,
G_\Delta)$ from Eq.\ref{corrxy4}.

In order to improve the statistics we have taken 500 randomly
chosen ways to divide the total season into two disjunct sets of
$N-1$ match days. We take care that match days k and k+N-1 (having the same matches just with an exchange of home- and away-teams) are not in the same set. In order to avoid any additional bias we have to
correct for the home advantage. In previous work \cite{Heuer09} we have
shown that for the Bundesliga the home advantage can be described
by a constant shift in the average goal difference which basically
does {\it not} depend on the individual team. Thus, we have
determined the home advantage for $X$ and $\Delta G$ for each
season and subtracted the home advantage from the individual
values.

In principle one may expect that the team strength and other properties (measured via the performance indicator $X$) of a team somewhat vary during the course of a season although the variation turns out to be quite small \cite{Heuer09}. Formally, the team strength is defined as the average team strength during the season. The residual variations of the team strength would show up as additional random effects, e.g., via the term $r(X,X)$.

\subsection{Estimation of statistical uncertainties}

A crucial aspect is the determination of the statistical
uncertainties of the different results, presented in this work. To obtain a reliable estimation
of these uncertainties, we have generated synthetic data which have the same statistical properties as the original data.
Specifically, we have drawn for each team a value for the
team strength and for the performance indicator $X$ from a
Gaussian distribution with the corresponding variances $\sigma^2(\Delta G)$ and $\sigma^2(X)$. Their respective choice is correlated in order to take into account the correct value of $r_{ideal}$. Subsequently the appropriate random contributions are added as determined by the variances $\epsilon^2(\Delta G)$ and $\epsilon^2(X)$. The estimation of these variances from the underlying data has been, e.g., described in Ref.\cite{Heuer14}. Repeating the same analysis with these synthetic data directly allows one to estimate the statistical uncertainties.

\section{Data description}
We choose different performance indicators for our analysis. From
www.kicker.de we obtained the {\it goals} and the {\it chances of goals} for
all matches from the season 2005/06 to the season 2016/17 for the
German soccer Bundesliga. Furthermore, from IMPECT \cite{internet5} we obtained
for the two seasons 2015/16 and 2016/17 the so-called {\it packing rate} and the
{\it impect rate}. Furthermore we include data about the number of
{\it successful passes} (for the seasons 2009/10 and 2010/11;
www.kicker.de) which are already published in a preliminary way
\cite{heuer_buch} and are reanalysed in the same framework as the
new data. Finally, the {\it expected goals (xGoals)} were obtained from
www.understat.com for the two seasons 2015/16 and 2016/17.

For  the definition of the packing and the impect rate all passes
and dribbles are analysed during a match; see also \cite{internet4}. The specific notations are chosen to be consistent with the previous notation of these observables. At a given time, $N_o$
denotes the number of players from the opponent standing closer to
the opponents goal than the ball. Whenever a pass (or dribble) is
successful, i.e the ball is still controlled by the ball
possessing team, one compares $N_o$(before) and $N_o$(after). If
$N_o$(before) $>$ $N_o$(after), the pass (or dribble) has bypassed
$N_o$(before) - $N_o$(after) players. This value is defined as the
packing rate for this specific event. The packing rate of a
match for a team is the sum over all individual packing rates. As
a consequence of this definition, safe passes to the back of the
field do not contribute. The impect rate is defined in
analogy to the packing rate. However, one has the additional
constraint that only the 6 players of the opponent, closest to
their goal, are taken into account. Thus, the impect rate
concentrates on events which, on average, are related to offensive
actions.

The chances for goals are defined by specific
criteria, consistently formulated by kicker.
Finally, the expected goals xgoals result from weighting each shot by
an empirical probability expressing the likelihood that a goal is scored
from this position on the pitch \cite{MacDonald12}.

For the performance indicator $X$ we mainly choose the difference
between the performance indicator of a given team and the
performance indicator for the opponents in the considered matches. A simple example is the goal difference, where
the number of goals of the opponents is subtracted from the own
goals. Typically, these difference values reflect best the
strength of a team since they contain both the offensive and
defensive aspects. For the example of goals we show that just taking the goals or the goals of the opponent significantly reduces the information content about the team strength.

\section{Results}
\subsection{General statistics}

\begin{table}[h!]
 \caption{The average values per team and match and
the home advantage (absolute and relative) for the different
performance indicators. If not mentioned otherwise the analysis is
based on the two seasons 2015/16 and 2016/17. The 12 year
interval, additionally analysed for the goals and the chances for
goals, are based on the seasons 2005/06 - 2016/17. The final three
performance indicators are related to passes. } \label{tab1}
      \begin{tabular}{llll}
        \hline
           & Average  & absolute home adv.    & relative home adv.\\ \hline
        Goals & 1.35  & 0.18   & 13 \%\\
        Goals (12 yrs.) & 1.43  & 0.18   & 13 \%\\
        xGoals & 1.39 & 0.16   & 12 \%\\
        Chances for goals & 5.6  & 0.7   & 13 \%\\
        Chances for goals (12 yrs.) & 5.6  & 0.7   & 13 \%\\
        \hline
        Passes & 294  & +13   & 4  \%\\
        Packing & 283 & +4 & 1 \%\\
        Impect & 40 & +1 & 3 \%\\\hline

      \end{tabular}
\end{table}

In Table 1  the average results as well as the home advantage
(absolut and relative) are listed for the different performance indicators,
discussed in this work. All performance indicators display a home
advantage. Remarkably, for the performance indicators, related to
passes, the home-away asymmetry is much smaller as compared to the
performance indicators related to scoring goals. Furthermore, the
huge difference in absolute numbers of the packing rate as
compared to the impect rate clearly indicates that most successful
passes, contributing to the packing rate, do not involve taking
out the defensive players (i.e., the last 6 players) of the
opponent.

\subsection{Information content about team strength}

\begin{table}[h!]
\caption{The relevant correlations coefficients,
characterizing the information content about the team strength.
The data are based on the 12 seasons 2005/06 -
2016/17.}\label{tab2}
      \begin{tabular}{llll}
        \hline
         X  & $r_{ideal}(X,G_\Delta )$  & $[r(X,X)]^{1/2}$    & $A_X$\\ \hline
      Goals$_\Delta$ & 1.00 $\pm$ 0.02  & 0.82 $\pm$ 0.01 & 0.82 $\pm$ 0.02\\
    Goals$_+$  & 0.94 $\pm$ 0.02 & 0.79 $\pm$ 0.01  & 0.74 $\pm$ 0.02\\
    Goals$_-$ & -0.91 $\pm$ 0.04  & 0.70 $\pm$ 0.02 & -0.64 $\pm$ 0.04\\
    Points  & 1.00 $\pm$ 0.02 & 0.76 $\pm$ 0.01  & 0.76 $\pm$ 0.02         \\\hline
      \end{tabular}
\end{table}

In Table \ref{tab2} we start by studying standard performance indicators, related to goals and points. By construction, the ideal
correlation of a performance indicator with itself is 1.00 since
it expresses the identical property of a team. This naturally
explains why $r_{ideal} = 1.0$ for the goal difference. The
reduction factor of 0.82 reflects the random contributions of scoring
goals, resulting from the finite information. When restricting oneself to the goals of a team, the ideal
correlation is smaller than unity (0.94). This reflects the fact
that the defensive strength is no longer taken into account so
that information about an important ingredient of the team
strength is missing. Interestingly, when taking the goals of the
opponent as a performance indicator the ideal correlation further
decreases. This suggests that the offensive strength contributes
 stronger to the overall team strength than the defensive
strength. Finally, also the points
contain full knowledge about the team strength. However, the
random contributions are somewhat larger than for the goal
difference (smaller value of $r(X,X)^{1/2}$). This is not
surprising since a 6:0 and 1:0 are counted identically for points
whereas the goal difference takes into account that additional
information.

\begin{table}[h!]
\caption{The relevant correlations coefficients, characterizing
the information content about the team strength. Note that for all
performance indicators the differences (e.g. the goal differences)
are compared. If not mentioned otherwise two seasons are taken
into account. } \label{tab3}
      \begin{tabular}{llll}
        \hline
         X  & $r_{ideal}(X,G_\Delta)$  & $[r(X,X)]^{1/2}$    & $A_X$\\ \hline
     Goals$_\Delta$  & 1.00 $\pm$ 0.04  & 0.85 $\pm$ 0.02 & 0.85 $\pm$ 0.04\\
      Goals$_\Delta$  (12 yrs.) & 1.00 $\pm$ 0.02  & 0.82 $\pm$ 0.01 & 0.82 $\pm$ 0.02\\
        xGoals$_\Delta$  & 0.99 $\pm$ 0.04 & 0.89 $\pm$ 0.02   & 0.88 $\pm$ 0.04\\
        Chances for goals$_\Delta$  & 0.97 $\pm$ 0.03  & 0.92 $\pm$ 0.01   & 0.90 $\pm$ 0.03 \\
        Chances for goals$_\Delta$ (12 yrs.)  & 1.00 $\pm$ 0.01  & 0.91  $\pm$ 0.005  & 0.90 $\pm$ 0.01  \\\hline
       Passes$_\Delta$ & 0.88 $\pm$ 0.06  & 0.94 $\pm$ 0.01 & 0.83$\pm$ 0.06\\
       Packing$_\Delta$ & 0.96 $\pm$ 0.03 & 0.98 $\pm$ 0.005 & 0.94 $\pm$ 0.03\\
      Impect$_\Delta$ & 0.89 $\pm$ 0.07 & 0.83 $\pm$ 0.02 & 0.74 $\pm$ 0.06\\
       \hline
      \end{tabular}
\end{table}

Next we analyse different performance indicator within this approach. The results are listed in Table \ref{tab3}.  We start by discussing them one by one. We always consider
the value, obtained by a given team minus the value, obtained by
the opponents of that team in analogy to the goal difference, already discussed above.

Comparing the results for the goal difference during the last 2
years with the last 12 years, the noise contribution is larger
when taking 12 years (0.85 vs. 0.82). A more detailed analysis shows that during
the last 6 years the variance of the goal difference was
significantly larger than during the 6 years before. Basically, it
is due to Bayern M\"unchen, having gained a very positive goal
difference every year. For example in the season 2016/17 Bayern M\"unchen had
a goal difference of +67 whereas the second best value was +32.
Naturally, as expressed in Eq.\ref{corrxx}, an increase of the
variance reduces the noise effects.

For the expected goals (xGoals) the ideal correlation is again  basically
unity (0.99) whereas the noise contribution is significantly
smaller than for the goals themselves.  Thus, rather than
weighting each shot by +1 or 0 (goal or not goal) the information
content of that goal-related performance indicator increases by
using a distribution of weighting factors. Thus, this expected
goal variable can additionally distinguish between teams which on
average score from better or worse
 positions on the pitch.

For the chances for goals  the information content of
 about the team strength is comparable to the expected goals within the statistical uncertainties $(A_X = 0.90$ vs. 0.88).  It may come as a
surprise that $r_{ideal}$ is very close to unity in particular
when analysing the 12 years data set, displaying smaller
statistical uncertainties. This directly implies that the
efficiency of a team to score a goal, when having a chance for a
goal, is nearly constant. Indeed, this has been explicitly shown
in Ref.\cite{Heuer14}.  Thus, from a statistical point of view,
scoring of goals is basically identical to tossing a dice with
four sides, once a chance for goals has been generated. This
naturally explains why it is by far better to take the previous
chances for goals rather than the previous goals to forecast the
future goals of a team ($A_X = 0.90$ vs. 0.82 for the longer
12-year period).

Next we discuss the pass-related indicators. The ideal correlation
of the number of passes  with the team strength is 0.89,
indicating a strong but by far not perfect correlation. In
contrast, the reduction factor due to noise is closer to unity
than for any goal-related indicator (0.94). This expresses the fact
that the number of passes per match is large so that random
effects are automatically reduced. Finally,  the overall
information content ends up to be as low as $A_X =0.83$ which is
close to the results for the number of goals.  We mention in
passing that similar statistical properties are obtained for
observables such as the number of different contacts with the
ball, reflecting the ball possession.

Most remarkably, the packing rate is, firstly, highly correlated
with the team strength (0.96) and, secondly, displays a neglible
reduction of the information content due to  noise (0.98). As a
consequence, the value of $A_X = 0.94$ is larger than all the
values, discussed so far. Thus, from a practical point of view the
definition of the packing rate results in a highly versatile
performance indicator. The observation that the ideal correlation
is larger than for the number of successful passes clearly
indicates that the weighting of a pass with the impact of the
passes is a major improvement to highlight the strength of a team.
Stated differently, the ability to play smart successful passes is
a key ingredient to finally generate chances for goals as a
prerequisite for scoring goals. Also the noise is significantly
smaller than for the number of successful passes (0.98 vs. 0.94). This
may come as a surprise since both the packing rate and the number of successful
passes have similar absolute numbers per match (see Table 1). This suggests
that there is a significant match-to-match variation of the number of passes,
e.g. as a consequence of different tactical measures. In contrast, the packing rate is
a more direct measure of the quality of the team and therefore shows up
similarly in all matches of that team.

The general observations, reported for the packing rate, fit very well to the specific observation that
during the two seasons, analysed in this work, Bayern M\"unchen
displayed a higher packing rate in {\it all} matches as compared
to the opponent. In contrast, in some matches Bayern M\"unchen did
not win. This reflects the fact that the random effects for goals are higher than for
the packing rate.

Naively, one might expect that the impect rate is even more
relevant since it reflects match actions closer to the goal of the
opponent. However, the statistical analysis reveals that the ideal
correlation is smaller (0.89 vs. 0.96).  This suggests that the
strength of a team is better reflected by the quality of the
passes without the restriction to analyse only the region closer
to the goal. Furthermore, also the noise contribution is much
larger than for the packing rate (0.83 vs. 0.98 for the reduction
factor). The latter is to be expected because the actual values of
the impect rate (see Table \ref{tab1}) are much smaller than the
packing rate (40 vs. 283). Taking all aspects together, the impect
rate has no relevance for the purpose of forecasting.

\subsection{Single-match correlations}

So far we have analysed how well the information from previous
matches can help to estimate future results. A quantitative
measure was $A_X$ for the performance indicator $X$.  A different
but also relevant question deals with correlations within a
single match. Which performance indicator $X_\Delta$, recorded in a
match, is best in explaining the goal difference of a single match
of team A vs. team B? Both quantities for that specific match are
denoted $x_{AB}$ and $g_{AB}$, respectively.

\begin{table}[h!]
 \caption{Correlations within single matches. See main
text for closer description. }\label{tab4}
      \begin{tabular}{lll}
        \hline
         $x$  & $r(x_{AB}, g_{AB}) $ & $r(x_{AB} - \langle x_{AB} \rangle, g_{AB} - \langle g_{AB} \rangle) $ \\ \hline
        Packing & 0.36 $\pm 0.04$  & -0.02 $\pm 0.04$   \\
       Impect& 0.64 $\pm 0.02$ & 0.58  $\pm 0.02$  \\
      Chances for goals & 0.65 $\pm 0.02$ & 0.52 $\pm 0.03$ \\
        xGoals & 0.62 $\pm 0.02$ &  0.52 $\pm 0.03$ \\\hline
      \end{tabular}
\end{table}

First, we analyse the Pearson correlation
$r(x_{AB},g_{AB})$. The values are listed in Table \ref{tab4}. All
performance indicators display a positive correlation.
The correlation for the packing rate is
significantly smaller than for the other performance indicators, namely the  impect rate,
chances for goals and expected goals (correlation coefficient
above 0.6). This shows, that the observation for the World Cup
match Germany vs. Brazil, mentioned above, has a systematic
background: the impect rate is indeed a strong indicator of the actual result.

Partly, this result just reflects the ideal correlation, analysed
before. If in the previous matches team A on average had
generated, e.g., many more chances for goals than team B, it is
likely that team A will win against team B in that specific match.
This contributes to the positive correlation of the chances for
goals and the actual result and thus does not convey much new information.

However, new information is obtained, if one takes out this expected
correlation and rather compares the respective differences to the
expected values, i.e. $x_{AB} - \langle x_{AB} \rangle$ to $
g_{AB} - \langle g_{AB} \rangle$. For example one might expect
that a result 5:1 of two teams with similar team strength should
go along with an unbalanced number of chances for goals. Thus, one
explicitly probes whether the statistical fluctuations of the
number of goals is directly related to fluctuations in the
performance indicator $X$. Strong correlations would indicate the a particular
large value of that performance indicator in a match is a prerequisite for
scoring goals.

In a first step, we predict the expected outcome from the
respective team strengths via $g_{AB} = [(N-1)/N] (S_A - S_B)$. The prefactor guarantees that after averaging over all opponents B the average goal difference of team A
is exactly $S_A$, as it should be by definition. Again we
subtract any effects, resulting from the home advantage (see
above). Following the discussion in Ref.\cite{Heuer10} the
expected difference $\langle \Delta x_{AB} \rangle$ can be written
as $c (X_{\Delta,A} - X_{\Delta,B})$, denoting
the difference performance indicator  $X$, averaged over the
remaining $2N-1$ matches of the season of both teams,
respectively.  The proportionality constant $c$ reflects the
regression towards the mean and can, in principle, be estimated
from the term $r(X,X)$, discussed above. In practice, we have
performed a straightforward regression analysis in order to
estimate $c$ which, in the limit of large $N$, does approach
unity. In the second step the correlation coefficients can be
directly determined. The results are listed in Table \ref{tab4}. Naturally, for the choice $x_{AB} = g_{AB}$ we would obtain a correlation coefficient of strictly one. Intuitively, this reflects the trivial 100 \% correlation between goals and goals in the same match.

Interestingly, it turns out that a larger or smaller packing rate
as compared to the expectation has no impact on the actual result
relative to the expected one. This may come as a surprise since one might expect that an increased packing rate relative to expectation may generate a larger number of chances of goals (because of strong relation between packing rate and team strength) and thus, finally, also more goals. For a closer understanding one needs to take into account that the noise effect for the packing rate is very small ($1 - 0.98 \ll 1$ ; see Table 3). As a consequence the relevant variable $x_{AB} - \langle x_{AB}\rangle $ displays only minor fluctuations. These fluctuations are not large enough to explain any variation from the expected outcome of the match.

In contrast, strongly positive correlations are
observed for variables which are directly expressing attacking
actions. For example, the chances for goals display a significant correlation of 0.52. This can be easily rationalized, based on the causal relation between chances for goals and resulting goals. We use
the following simple assumptions: (i) All teams have the same
team strength. This assumption is reasonable because
heterogeneities are already taken care by subtracting the average
values. (ii) Chances for goals and goals are distributed according
to a Poissonian distribution. (iii) No correlations between the
goals of teams A and B are present, i.e. one has independent
Poissonian processes. The same holds for the chances for goals
(iv) Scoring a goal after achieving a chance for a goal is a
random process with probability $0.25$ \cite{Heuer14}. In this final assumption a causal relation between chances for goals and actual number of goals enters. The average
total number of chances for goals is denoted $c_{tot}$, the
average total number of goals is denoted $g_{tot}  = 0.25 c_{tot}$. Then one
can write $ \langle \Delta x \Delta g \rangle / [\sqrt{\langle
\Delta x \rangle^2 }\sqrt{\langle \Delta g \rangle^2 }] = 0.25
\langle \Delta x^2 \rangle /[\sqrt{\langle \Delta x^2 \rangle
}\sqrt{ \langle \Delta g^2 \rangle }] =
0.25c_{tot}/[\sqrt{c_{tot}}\sqrt{g_{tot}}] = 0.50 $ in excellent
agreement with the actually observed value of 0.52.

Also the performance indicator xGoals displays the same correlation as the chances for goals.  As already discussed in the
context of Table \ref{tab3},  this shows again that the statistics of
xGoals is strongly connected with the chances for goals.

Surprisingly, also the impect rate displays a high correlation of similar magnitude (0.58) and thus behaves very differently than the packing rate. As already discussed in the context of Tab.3, the impect rate is characterized by a much higher noise contribution (0.83) than  the packing rate (0.98). Thus, we just have to invert the arguments from above, used for the packing rate. The resulting strong fluctuations of the impact rate give rise to matches where team A has many more excellent passes close to the goal than expected. This may naturally increase the number of goals as expressed by  a large value of $r(x_{AB}, g_{AB}) $.

\section{Discussion}

We have presented a systematic  approach to discuss several
distinct aspects of performance indicators and applied this to the
specific case of soccer. In particular due to the availability of
tracking data a large number of performance indicators can be defined. They may serve
different purposes. However, when coming to the gold standard of
forecasting soccer matches or, more precisely, estimating the team
strength, the performance indicators behave very differently. The packing rate turns out to be the most informative
performance indicator.

Please note that this result could have been also obtained by straightforward determination of the Pearson correlation between the respective performance indicator in the first half of the season and the goal difference in the second half. The new aspect of the present approach is an improved understanding of why some performance indicators are better suited to predict the team strength than others. This is achieved by decomposing the overall forecasting quality into two contributions. For the example of the packing rate it turns out that, firstly, it has
a very high correlation with the team strength when excluding
random effects (0.96) and, secondly, also displays very little random effects (0.98). Due to appropriate normalization these values can be directly compared with the optimum values of unity (perfect correlation with team strength and no random effects, respectively).
Further informative performance indicators are the chances for goals and the
xGoals. They display a significantly improved information context as compared to the actual number of
goals. This can be related to the reduced random contributions.

Although these results may have implications for optimizing the forecasting of soccer matches, the main motivation of this work is to set a framework for the identification of performance indicators which are most closely related to the final aim of a soccer team, namely to score goals. By definition the ability of scoring goals (and avoiding goals by the opponent) is directly related to its team strength. The results in Table 3, obtained from correlating two different sets of matches, reflect the value of $r_{ideal}$. It exactly expresses to which degree a performance indicator is relevant for the team strength. Naturally, for performance indicators, which are causally related to scoring goals (goals, chances for goals, expected goals) a high value of $r_{ideal}$ is expected. However, it may come as a surprise that also the packing rate displays such a high value of $r_{ideal} ( 0.96 \pm 0.03)$.  This expresses the fact that the generation of chances for goals via intelligent passes on the pitch is a core competence of a successful team.
In contrast, the mere number of  passes displays a significantly smaller value of $r_{ideal}$ than the packing rate and is thus less correlated with the team strength. This reflects the fact that via some tactical measures it may be quite straightforward to increase the number of passes but not so much the quality of the passes. Rather any increase of the packing rate requires a fundamental improvement of the team strength which is beyond simple tactical measures.

Naturally, for forecasting processes one may try to use the information from more than one performance indicator to improve the estimation of the team strengths in a league and thus of future match results. Actually, the relevant statistical background has been, e.g., discussed in Ref.\cite{Heuer14}. Of course, this is particularly useful if the different performance indicators contain somewhat different pieces of information. Thus, it might be expected that the combination of the packing rate with, e.g., the number of chances for goals (or the expected goals) may be a reasonable guess. However, this analysis is beyond the scope of the present work.

The setup, chosen in this analysis, can be directly used to quantify the forecasting quality after half of the matches of a season are played. More generally, one can ask the question about the quality of the estimation of the team strength after $M$ match days ($1 \le M \le 2N-3$). Whereas $r_{ideal}$ does, by construction, not depend on $M$, the noise contribution naturally increases for decreasing $M$. Following a straightforward mathematical analysis one can show that this increase of noise effects is more pronounced for performance indicators with stronger noise contributions, i.e. smaller values of $r(X,X)$.  As a consequence, the superiority of the packing rate would be even more pronounced if the team strength is estimated after only a few match days. Furthermore, for small values of $M$ it may be also useful to incorporate information from previous seasons. Naturally, due to changes of the team strength from season to season the ideal correlation with the team strength will not be so large. However, since that performance indicator contains the information from the whole season, this deficiency is counterbalanced by a relatively small noise contribution. Thus, in practice, the information from the previous season may still be relevant for forecasting purposes after just a few match days.

It may be interesting to apply the general statistical framework
also to other types of sports, trying to identify the
underlying secrets of successful teams or to other situations where
random and systematic effects are present at the same time. Furthermore, within the present framework it can be easily checked for new performance indicators, obtained from tracking data, to which degree the team strength is reflected.

\vspace{1cm}

{\bf Acknowledgement}

We gratefully acknowledge S. Reinartz and L. Keppler for supplying the data about the packing rate and the impect rate and for helpful discussions. Understat is acknowledged for supplying the xGoals data. Finally, I would like to thank O. Rubner for technical support and helpful discussions.

\vspace{1cm}



\end{document}